# A Linear-Time and Space Algorithm for Optimal Traffic Signal Durations at an Intersection

Sameh Samra, Ahmed El-Mahdy
Computer Science Engineering Department, Egypt-Japan University of Science and Technology (E-JUST), Alexandria, Egypt, {sameh.abosamra@ejust.edu.eg, ahmed.elmahdy@ejust.edu.eg}

Yasutaka Wada
Graduate School of Information Systems, The University of Electro-Communications, Tokyo, Japan, {wada@is.uec.ac.jp}

*Abstract*— **Finding an optimal solution of signal traffic control durations is a computationally intensive task. It is typically $O(T^3)$ in time, and $O(T^2)$ in space, where $T$ is the length of the control interval in discrete time steps. In this paper, we propose a linear time and space algorithm for the same problem. The algorithm provides for an efficient dynamic programming formulation of the state space, the prunes non-optimal states, early on. The paper proves the correctness of the algorithm and provides an initial experimental validation.**

## 1. INTRODUCTION

As population in large cities increases, the traffic problem becomes more serious. A classical traffic control problem is finding optimal traffic phase durations for an intersection, to ensure smooth and safe passage of vehicles.

A classical algorithm for handling this problem is the Controlled Optimization of an Intersection (COP) algorithm [1]. The algorithm utilises dynamic programming to find the optimal durations for a changing real-time traffic patterns. However, the algorithm achives a time complexity of $O(T^3)$ and space complexity of $O(T^2)$, where $T$ is the length of the control interval, in discrete time steps.

In this paper, we propose a novel algorithm that tackles the same problem, while reducing the time and space complexities into $O(T)$. The algorithm utilizesan efficient dynamic programming formulation of the traffic problem. The formulation exploits a phase, rather than interval time, to eliminate many unnecessary computations, reducing both computation time and memory space requirements. As a result, the proposed algorithm achieves more than 2700 times speedup (determined experimentally) against the original COP algorithm for $T$ =1024. We choose $T$ to be the nearest powers-of-two to the maximum typical traffic time-horizon prediction of 15 min (900 seconds), with one second unit [2], [3].

The paper is organized as follows: Section 2 discusses related work. Section 3 introduces the traffic control problem; Section 4 introduces the new proposed algorithm; Section 5 validates the proposed algorithm. Section 6 presents and discusses the experimental results, and finally Section 7 concludes the paper

## 2. RELATED WORK

There are many strategies to solve traffic control problem efficiently. Authors in [4] survey these strategies. These include dynamic programming, neural networks [5], multi-agent systems [6], petri nets [7], genetic algorithm [8] and fuzzy control [9].

The strategies can be characterized by the number of considered traffic intersections, and how they model the traffic demand and response [10]. A control strategy can produce a time plan for a single or multiple interactions. The former are called 'isolated' strategies, and the latter are called 'coordinated' strategies. The traffic demand can be statically determined using (off-line) historical demands, and is used to produce 'fixed-time' traffic-control response strategies. Alternatively, the demand can be dynamically determined using real-time measurements to produce adaptive traffic response strategies. Combining these two characteristics, we get four possible categories of traffic control strategies.

Dynamic programming is used, generally, to solve optimization problems including the traffic control. There are many traffic control strategies that use dynamic programming; OPAC, PRODYN and UTOPIA are examples of these strategies. However, in their current version, they generally use approximate strategies [11] to decrease the computation complexity. RHODES is an exception, as it uses the dynamic programming algorithm COP to get an intermediate solution for each single intersection before generating the total solution for all intersections, trading computation speed for control accuracy.

ALLONS-D [12] and ADPAS [13] are another algorithms that make use of dynamic programming methodology. They use decision trees to get the optimal solution.However the time and space complexities of the algorithm are exponential, even with tree pruning.

The closest algorithm to our work is that proposed by Fang in her PhD thesis [14]. Her algorithm uses a time-oriented dynamic programming approach. However the algorithm is designedfor the specific case of three-phase dual-intersections,for two close intersections on a freeway (diamond interchange). The generalization is left for future research. Her implementation is based on selecting one phase as a start phase. Fang's algorithm and our proposed algorithm are using the same concept for selecting dynamic programming states, but our algorithm is general for any number of phases and studies all phases to determine the start one. Also Fang's algorithm does not consider clearance interval and minimum green time that are typically required in real-world traffic control system such as RHODES.

Due to the high time and space complexities of dynamic programming strategies, approximate or adaptive dynamic programming (ADP) strategies appeared and used to manage multiple intersections. Chen in [15] uses linear function approximation to reduce computations through transitions between states, and reinforcement learning is used to enhance the approximation. ADHDP strategy [16] makes use of reinforcement learning and dynamic programming to produce a near optimal solution for multi-intersections.

## 3. THE TRAFFIC CONTROL PROBLEM

The road intersection can be considered as a limited resource. The traffic control problem at one intersection is how to assign time for each traffic-flow direction to optimise a performance metric while maintaining a safe passage of cars, over a time horizon $T$. Typical metrics include total number of stops, waiting time, and queue lengths.

Figure 1 shows an example for a traffic intersection. The intersection consists of two crossing roads with eight possible directions, number from 1 to 8. The combinations of non-conflicting directionsare called *phases*; for example, directions 2 and 6 construct a phase, as we can give the right-of-way to these two directions in the same time, without breaking safety. Generally, we refer to the set of all possible phases a $P$ and number of phases as $|P|$.

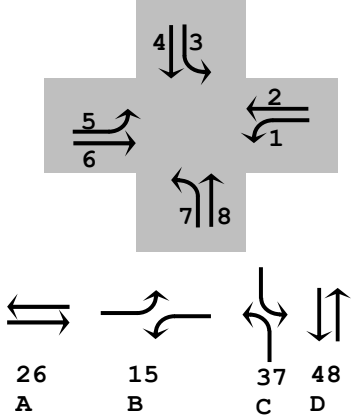

**Figure 1: Example Phases at a Single Traffic Intersection**

A solution of the traffic control problem is a sequence of phases with a time duration assigned to each phase to maximize a certain performance parameter, while taking into consideration non-contradicting phases and assigning a minium duration, $\gamma$, for each phase.This sequence is called the signal time-plan. Table 1 shows one example of the signal time-plan after assigning symbols to phases such that directions 2 and 6 are phase 'A'; directions 1 and 5 are phase 'B'; directions 3 and 7 are phase 'C'; 4 and 8 are phase 'D'.

The solution depends on the input taffic flows. The assumption here is that future can be predicted using current and historial data.Therefore, the input to the algorithm is number of vehicles arrive to each phase at specific future time.

**Table 1: Exemplar Signal Time-Plan**

| Phase | A | B | C | D | B | D |
|---|---|---|---|---|---|---|
| Duration(sec) | 3 | 5 | 4 | 6 | 3 | 5 |

More formally the optimisation problem can be formulated as: Define a timing plan, $\Omega$ to be the sequence $<(p_0, t_0),(p_1, t_1),…,(p_i, t_i)>$ such that:
- $p_i \in P$,
- $0 \leq i \leq T$,
- $t_i \geq \gamma$
- $\sum_i t_i = T$.

Also define $V(\Omega)$ to be the cost function of using the timing plan $\Omega$.

The optimisation can be stated as:

$$\text{Minimize}_k \ V(\Omega_k)$$

Where we seek of find the timing plan $\Omega_k$ that gives the minimum cost function $V()$

## 4. PROPOSED ALGORITHM

We use dynamic programming $(p_i, t_i)$ formulation to find an optimal timing plan, $\Omega$. For sake of simplicity, we start by setting the clearing duration r to one and minimum green time to one, we then generalize.

The time plan can be expressed as the sequence $\Psi(p_i,i)=<(p_0, 1),(p_1, 2), … , (p_i, i)>$, where $p_i$ is a traffic phase, including the clearance phase, where no traffic flows. In other words we define for each time unit, the active phase.

Define $\Psi(p_i, i)$ to be the optimal solutions at time $i$ ending at all possible phases, $p_i$; then the optimal solution at $i+1$ and a specific phases, $p_{i+1}$, can be computed as:

$$\Psi(p_{i+1},i+1) = \text{Min}_\rho \ V(\Psi(\rho,i) \cdot (p_{i+1},i+1))$$

$$\rho \in \begin{cases} \{0, p_{i+1}\} & p_{i+1} \neq 0 \\ P & p_{i+1} = 0 \end{cases}$$

Where $\cdot$ is the concatenation operator; the value of $\rho$ is constraint so as to prevent sudden change of traffic flows without a clearance interval.

We also define the cost function to be decomposable such that:

$V(\Psi \cdot (p_i,i)) = V(\Psi) \cdot V(p_i,i)$

The stages of dynamic programming are the time steps while each stages contains all possible states which describe the signal state of the traffic. The idea behind our algorithm is dealing with the green time units as a limited resource; it assigns current time unit to each available state to get the minimum cost function $V()$ for each state. To illustrate the core algorithm, we consider three phases (without loss of generality) as depicted in Figure 2. The phases are labeled $A$, $B$, and $C$. A state represents a current active phase of the traffic intersection, except for the clearance interval state, '0', where it represents a no-active phase. The arrows between states show the dependency between states and according to our objective function of dynamic programming. For sure, we cannot start with clearance state, so we have not any arrows get out from 0 state when $t=1$.

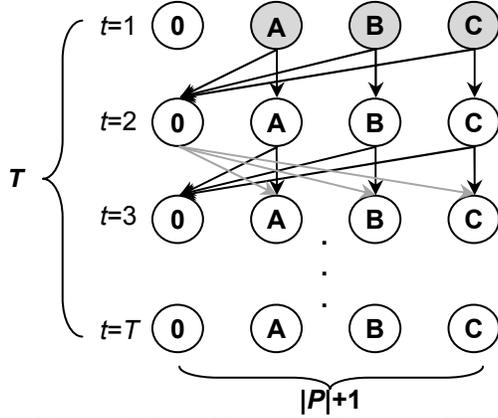

**Figure 2: State-Transition Diagram Where |P|=3, r=1 and γ=1**

To generalize our idea for any value of $r$ and $\gamma$, the number of states is computed as: $\gamma*|P|+r$; where ($r \leq \gamma$). These states can be grouped and labeled as shown in Table 2.

**Table 2: Dynamic Programming States**

| Group | Label |
|---|---|
| Stable clearance | 0 |
| Stable state | <phase_name> <br> <phase_name> := 'A'\| 'B'\| … |
| Unstable clearance states | <clearance_count> 0 <br> <clearance_count>:=1\|2\|3\|…\|r-1 |
| Unstable state | <count><phase_name> <br> <count> := 1 \| 2 \| 3 \| … \|γ-1 |

Stable states are the original states that represent live states at the traffic signal. However, the unstable states are virtual states created to keep the intermediate situations which are necessary for the next iteration. Table 3 extends the values of ρ in our objective function.

**Table 3: Groups of (ρ)**

| $P_{i+1}$ | ρ belong to |
|---|---|
| 10 | <phase name> |
| <clearance_count>0 | <clearance_count-1>0 |
| 0 | <r-1>0 |
| <count><phase_name> | <count-1><phase_name> |
| <phase name> | (γ-1)<phase_name>\|<phase_name> |
| 1<phase_name> | 0 |

The minimum green time imposes the constraint that at least a phase has to be active for γ units of time; the state is prefixed by a count of the number of time-units since the clearance phase and up to γ-1 time units; a prefix indicates that at the next time unit the phase cannot change; therefore a next state would have the same phase name but with a prefix count increased by one (or removed if it reaches γ). A phase with no prefix can stay the same or change to the clearance phase at the next time unit.

Figure 3 shows the corresponding state-transition diagram when $|P|=3$, $r=2$, and $\gamma=3$. Gray states in the first level are the initial states, other states cannot be initial state because they express more advanced status of the system. The transitions between states follow Table 2, which shows the set of predecessor states for each state. The arrows reflect these dependences between states. The target algorithm searches for the optimal solution through a matrix of dimensions $T$ by $\gamma|P|+r$. This is $O(|P|)$ as $\gamma$ and $r$ are constants.

The Trace-Back procedure searches for the state with minimum cost in the last stage with the condition that the state must be a stable one. Then using the predecessor state, we get the optimal path through the 2 dimensional states until reaching the initial stage.

COP uses the dynamic programming to obtain the optimal sequence of phases. COP recursion uses phases as stages in dynamic programming. The algorithm consists of two procedures: Forward Recursion and Trace-Back procedures [1]. It is worth noting the following:

- COP algorithm time complexity is $O(T^3)$ for a constant number of phases [8]; we will further analyze the algorithm in Section VII.
- COP algorithm explores the design space in a phase-oriented manner; exploring for each phase all possible time durations. This imposes a particular (cyclic) phase order, and results in a possible (linear) increase in the number of stages; for instance, as a worst case, a reverse phase sequence for phases $C$, $B$, $A$, could result in tripling the number of explored phase cycles (e.g. $A$, $B$, **$C$**, $A$, **$B$**, $C$, **$A$**, where the underlined stages are the sought ones, the other stages would have a time deltas of zero).

## 5. VALIDATION OF PROPOSED ALGORITHM

The validation of our new algorithm has two parts: The first is the theoretical proof of dynamic programming formulations, we show how it gets the optimal solution; the second is the experimental validation by comparing the results with COP for a variety of different inputs.

Our dynamic programming algorithm decomposes the problem into interdependent sub-problems arranged into stages; the solution of each sub-problem depends on one or more sub-problems from a previous stages and the objective function contains only one recursive term, so it is serial monadic dynamic programming [17]). Each sub-problem is evaluated only once and its value is kept.

Our objective function computes $\Psi(p_{i+1}, i+1)$ for all integer value $i<T$ so from the trace back procedure:

$\Psi_{\text{Final}} = \text{Min } \{\Psi(p_T, T)\}$ where $p_T \in P+\{0\}$

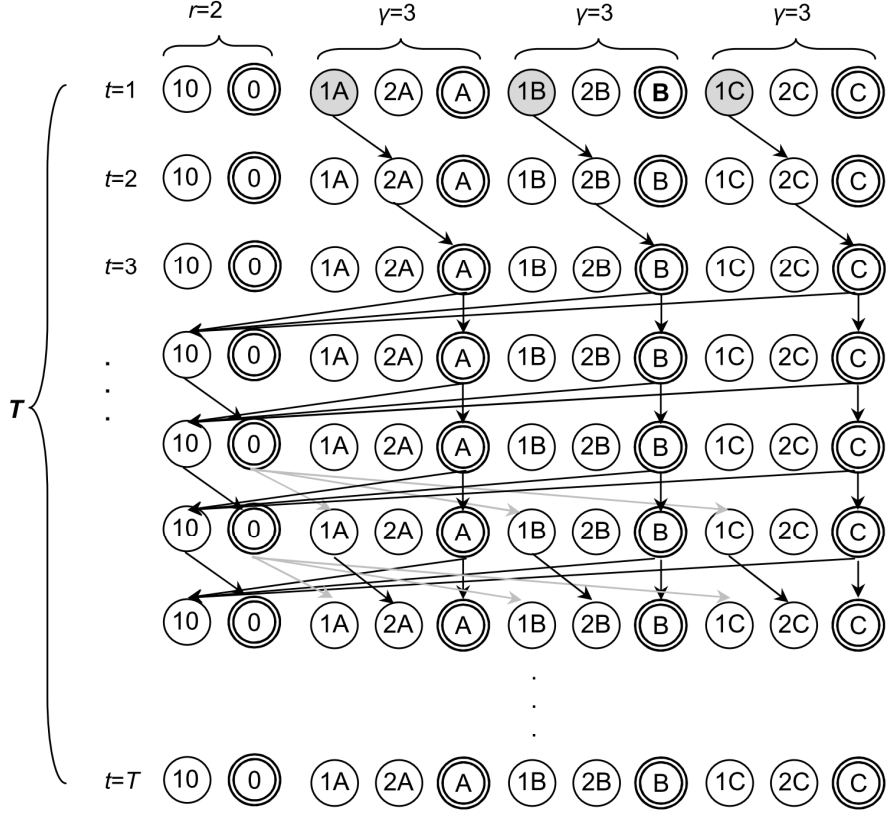

**Figure 3: State-Transition Diagram Where *r*=2 and *γ*=3**

Suppose $\Psi(p_j,j)$ is the sub-problem of finding the cost of the optimal traffic state sequence of length *j* that ends with the state $p_j$. The rationale here is that we can construct $\Psi(p_j,j)$ by only examining all optimal sequences of length *j*-1 each ending with a different state (with value function $\Psi(p_{j-1},j-1)$, for all (*j*-1) available states); we then choose the one that gives the minimum value when $p_j$ is the next state. This means that for each state we find the optimal sequence that brings us to this state by examining all possible previous states that leads to the current state.

We repeat the same procedure for all states until we reach past the final time step (*j* = *T*); this state is the final state. The value function of the final state is the minimum cost of all sequences of length T; therefore, it essentially chooses the optimal solution we are seeking.

We generate the optimal sequence of states by tracing back from the final state, choosing the previous state that generates a minimum value function. Figure 3 shows the forward recursion states and transitions between states. We keep the best previous state for each state in a table. At the last level, there is the optimal state; one of surrounded by double circles; we would start back tracking from that state.

### 5.1. Correctness Proof

This proves, by contradiction that it is not possible to construct an optimal sequence of length *j* that does not contain an optimal string of length *j*-1. Define $\Psi(p_j,j)$ to be the sequence: $<(p_0, 1),(p_1, 2), \ldots , (p_j, j)>$, of states (phases). Let || to be the sequence concatenation operator, defined as: $\Psi(p_i,i)$ || $\Psi'(p_j,j) = <(p_0, 1),(p_1, 2), \ldots , (p_i, i)>, <(p''_0, 1),(p''_1, 2), \ldots , (p''_j, j)>$.

Assume that it is possible to find the sequence $\Psi(p_j,j)$ such that the sub-sequence $\Psi'(p_{j-1},j-1)$ is not an optimal sequence, ending with state $(p_i, i)$. Designates*(*j*-1) to be an optimal sequence ending with $(p_i, i)^* =(p_{j-1}, j-1)$. Therefore, we can construct a new sequence $(p_j, j)^* =(p_{j-1}, j-1)^*|| (p_j, j)$ that has a lower cost than $(p_j, j)$, which is a contradiction. Therefore, the sequence $\Psi'(p_{j-1},j-1)$ has to be an optimal sequence ☐

### 5.2. Experimental Validation

The second part of the validation is experimentally done by comparing the generated optimal solutions with that of COP; we varied *T* to grow in according to the geometric sequence: 8,16,... ,4096. Repeating the arrival data used in [1]. Both algorithms generate the same optimal solution.

### 6. EXPERIMENTS AND RESULTS

We implemented the two algorithms in C. We evaluated the two algorithms on the same PC with the configuration of Intel Core2 Duo @ 2.10GHz with 2GB of RAM. The traffic control problem as formulated has the following parameters: number of phases, number of discrete-time steps and the traffic-load. For a single intersection, the discrete-time steps parameter is the limiting factor in performance. In our experiments, we used the same traffic parameters as the original

COP algorithm; we set phases to three and used the same traffic-load as in the paper (repeated to match with $T$, as in the validation section). We study the program execution time against varying the discrete-time steps, $T$. Figure 4 and 5 show the execution time in seconds for COP and the proposed algorithm respectively. Our algorithm clearly improves the execution time, reaching about 2700 times speedup when T = 1024. The growth rate is largely linear of the proposed algorithm and cubic for COP.

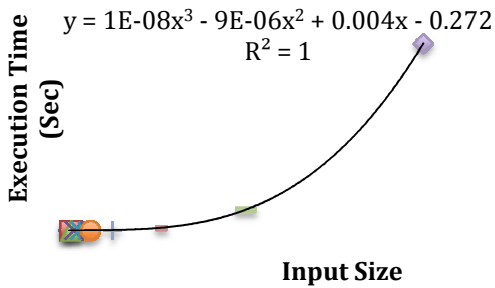

**Figure 4: COP Execution Time and/ Curve Fitting**

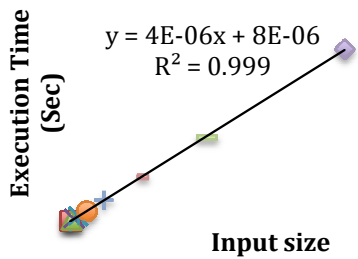

**Figure 5: The new algorithm Execution Time and Curve /Fitting**

The COP and the proposed algorithms space complexities are $O(|P|T(T-r))$ and $O((\gamma|P|+r)T)$ respectively, which are the sizes of the corresponding matrices. Table 4 summarizes the time and space complexities for both algorithms.

**Table 4: COP and New Algorithm Time and Space Complexities**

|  | Time Complexity | Space Complexity | $|P|$, $\gamma$ and $r$ are constants | |
|---|---|---|---|---|
|  |  |  | Time Complexity | Space Complexity |
| COP | $O(|P|^2T(T-r)(T-\gamma))$ | $O(|P|T(T-r))$ | $O(T^3)$ | $O(T^2)$ |
| Our algorithm | $O((\gamma|P|+r)T|P|)$ | $O((\gamma|P|+r)T)$ | $O(T)$ | $O(T)$ |

## 7. CONCLUSION AND FUTURE WORK

This paper presents a new dynamic programming algorithm to find an optimal solution of the traffic control problem at an intersection. The proposed algorithm has linear time and space complexities; it achieves an $O(T^2)$ performance, and $O(T)$ space enhancements, over the well-known COP algorithm. Consequently, it reduces computations power and the energy allowable for more scalable real-world applications. The algorithm improves time and space complexities by not abiding with strict (cyclic) phase order, and exploring the solution space in a time-oriented fashion, pruning many non-optimal states early on. The paper provides a correctness proof and an initial experimental validation.

Future research includes the integration of our algorithm with coordinated/traffic responsive strategy such as RHODES [11], and building parallel version of it, to model multiple intersections. Moreover, we are currently exploring an optimization for enhancing the cost function computation time to $O(1)$ instead of $O(|P|)$ for each dynamic programming state.